\pdfoutput=1

\documentclass[namedreferences]{SolarPhysics}
\usepackage[optionalrh,solaenum,natbib]{spr-sola-addons} % For Solar Physics
\usepackage{graphicx}        % For eps figures, newer & more powerfull
\usepackage{amssymb}        % useful mathematical symbols
\usepackage{color}           % For color text: \color command
\usepackage{url}             % For breaking URLs easily trough lines
\usepackage{apjfonts}
\usepackage{rotating}
\newcommand{\rmd}{ {\ \mathrm d} }
\renewcommand{\vec}[1]{ {\mathbf #1} }

\newcommand{\mB}{\mathcal{B}}

\newcommand{\aap}{    {\it Astron. Astrophys.}}

\newcommand{\apj}{    {\it Astrophys. J.}}

\newcommand{\solphys}{{\it Solar Phys.}}

\newcommand{\ssr}{    {\it Space Sci. Rev.}}

\newcommand{\Eq}{{Equation}}

\newcommand{\Fig}{{Figure}}
\newcommand{\Figs}{{Figures}}

\graphicspath{{../artb/Figs/}} % define the path of all the Figures
\begin{document}
\begin{article}
%-------------------------------------------------
\begin{opening}
  
  \title{Preprocess the Photospheric Vector Magnetograms for NLFFF
    Extrapolation using a Potential Field Model and an Optimization
    Method}

  \author{Chaowei Jiang$^{1}$, Xueshang Feng$^{1}$}

  \date{}
  
  \runningauthor{Jiang and Feng}
  \runningtitle{Preprocessing of vector magnetograms}
  
  \institute{$^{1}$SIGMA Weather Group, State Key Laboratory for Space
    Weather, Center for Space Science and Applied Research, Chinese
    Academy of Sciences, Beijing 100190\\
    $^*$ The corresponding author email: cwjiang@spaceweather.ac.cn}
  \begin{abstract}
    Numerical reconstruction/extrapolation of coronal nonlinear
    force-free magnetic field (NLFFF) usually takes the photospheric
    vector magnetogram as input at the bottom boundary. Magnetic field
    observed at the photosphere, however, contains force which is in
    conflict with the fundamental assumption of the force-free model
    and measurement noise which is unfavorable for practical
    computation. Preprocessing of the raw magnetogram has been
    proposed by \citet{Wiegelmann2006b} to remove the force and noise
    for providing better input for NLFFF modeling. In this paper we
    develop a new code of magnetogram preprocessing which is
    consistent with our extrapolation method CESE--MHD--NLFFF
    \citep{Jiang2012apj1,Jiang2012apj}. Basing on a magnetic-splitting
    rule that a magnetic field can be split into a potential field
    part and a non-potential part, we split the magnetogram and deal
    with the two parts separately. Preprocessing of the magnetogram's
    potential part is based on a numerical potential field model, and
    the non-potential part is preprocessed using the similar
    optimization method of \citet{Wiegelmann2006b}. The code is
    applied to the {SDO}/HMI data and results show that the method can
    remove efficiently the force and noise and improve the quality of
    extrapolation.
  \end{abstract}
\end{opening}

\section{Introduction}
\label{sec:intro}

Magnetic field extrapolation is an important tool to study the
three-dimensional (3D) solar coronal magnetic field, which is
difficult to measure directly \citep{Sakurai1989, Aly1989, Amari1997,
  McClymont1997, Wiegelmann2008, DeRosa2009}. The models being used
most popularly for field extrapolation are the potential field model,
the linear force-free field model, and the nonlinear force-free field
(NLFFF) model. These models are all based on the same assumption that
the Lorentz force is self-balancing in the corona, but adopt different
simplifications of the current distribution. Among these models, the
NLFFF model is the most precise one for characterizing magnetic field
in the low corona, where there is significant and localized electric
current, especially in active regions.

Regarding the NLFFF extrapolation, it is routine to use the vector
magnetograms observed on the photosphere as input, at least in most of
the available extrapolation codes \citep[{\it e.g.},][]{Wheatland2000,
  Wiegelmann2004, Amari2006, Valori2007, Jiang2012apj,
  Jiang2012apj1,Inoue2011}\footnote{There are also some NLFFF models
  which use only the line-of-sight component of the photosphere field,
  along with constraints from other observed information like the EUV
  loops, filament channel, and X-ray sigmoid structure
  \citep[{\it e.g.},][]{Bobra2008,Su2009,Aschwanden2012ApJ}.}. This,
however, poses a basic problem (also a major headache) to the
force-free field modelers, because the magnetic field in the
photosphere is forced by the plasma significantly \citep{Metcalf1995},
which is in conflict with the fundamental assumption of
force-freeness.
% At the photosphere the plasma is much denser than up in the corona,
% and
From the photosphere to the corona, the magnetic field passes through a highly
stratified and inhomogeneous plasma environment with plasma $\beta$
varying abruptly from $> 1$ to $\ll 1$ \citep{Gary2001}; thus the
force-free condition cannot be fulfilled globally. By a study of the
observed chromospheric field in a sampled active region,
\citet{Metcalf1995} conclude that the magnetic field is not force-free
in the photosphere, but becomes force-free roughly $400$~km above the
photosphere. A recent statistical study by \citet{Liu2012ff} using a
large number of magnetograms from Huairou Solar Observing Station
gives similar conclusions.

This complication leads to the desire to use measurements of the
vector field in the force-free upper chromosphere instead. However,
the vector field is not as easily measured in the chromosphere as in
the photosphere. Even the chromospheric field is measured, it is still
problematic for extrapolation since the surface in which any
particular magnetically-sensitive line will form varies in time and
space, and in particular the height will be different along different
lines of sight. So one cannot assume that the vector field is given on
a plane or sphere at the bottom of the extrapolation volume as in the
case of the photospheric magnetograms. The practical use of the
chromospheric magnetograms as boundary conditions for extrapolations
is still to be explored.

As an alternative way to alleviate the problem, one can consider to
modify the photospheric magnetograms to simulate the force-free
chromospheric magnetograms, which is first suggested by
\citet{Wiegelmann2006b}. Since the interface between the photosphere
and the bottom of the force-free domain is rather thin (say, about
400~km), especially if compared with the spatial scale of the coronal
field (about tens of megameters), the basic field structures of the
chromosphere should be very similar to those of the photosphere,
except that (i) there must be some smoothness of the structures due to
the fast expansion of field from the high-$\beta$ to low-$\beta$
regions and (ii) the very fine magnetic elements are just closed
within this interface and thus show no signal in the force-free
domain. Thanks to these reasons, modifications that need to be made on
the photospheric field to mimic the force-free chromospheric field
ought to be not significant and can hopefully be made
within/around the error margins of the measurement.

The procedure of modifying a raw photospheric magnetogram to a
force-free chromospheric one is usually called `preprocessing'
\citep{Wiegelmann2006,Fuhrmann2007,Metcalf2008,Fuhrmann2011,Yamamoto2012}. To
guide the preprocessing, there are constraints that must be fulfilled
by the target magnetogram. On the boundary surface $S$ of an ideally
force-free field $\vec B$ in a volume $V$, the field satisfies the
following necessary conditions\footnote{The necessary conditions mean
  that even fulfilling these conditions, the magnetogram may still
  contain force; but magnetograms with these conditions fulfilled are
  certainly better input for NLFFF model than those not.}
\begin{eqnarray}
  \label{eq:m1}
  F_{x} = \int_{S}B_{x}B_{z}\rmd x\rmd y = 0,\ \ 
  F_{y} = \int_{S}B_{y}B_{z}\rmd x\rmd y = 0,\nonumber\\
  F_{z} = \int_{S}E_{B}\rmd x\rmd y = 0,\ \
  T_{x} = \int_{S}yE_{B}\rmd x\rmd y = 0,\ \
  T_{y} = \int_{S}xE_{B}\rmd x\rmd y = 0, \nonumber\\
  T_{z} = \int_{S}(yB_{x}B_{z}-xB_{y}B_{z}) \rmd x\rmd y = 0.
\end{eqnarray}
where $E_{B} = B_{x}^{2}+B_{y}^{2}-B_{z}^{2}$. These expressions are
derived from the volume integrals of the total magnetic force and
torque \citep{Aly1989, Sakurai1989,
  Tadesse2011PhD}
\begin{eqnarray}
  \label{eq:m11}
  \vec 0 = \int_{V} \vec j\times \vec B \rmd V = \int_{V} \nabla \cdot
  \vec T
  \rmd V = \int_{S}\vec T \rmd \vec S,
  \nonumber\\
  \vec 0 = \int_{V} \vec r\times (\vec j\times \vec B) \rmd V = 
  \int_{V} \nabla \cdot \vec T' \rmd V = \int_{S}\vec T'\rmd \vec S
\end{eqnarray}
where $\vec T$ is magnetic stress tensor,
\begin{equation}
  \vec T_{ij} = -\frac{\vec B^{2}}{2}\delta_{ij}+B_{i}B_{j}
\end{equation}
and $\vec T_{ij}'=\epsilon_{ikl}r_{k}\vec T_{lj}$. Generally the
surface integration has to be carried out over a closed volume, but in
preprocessing magnetograms for extrapolation of a computational cube,
the surface integrals of \Eq~(\ref{eq:m11}) is usually restricted
within the bottom magnetogram since the contribution from other (side
and top) boundaries is small and negligible, and in the following $S$
will represent only the area of magnetograms. With this assumption,
\Eq~(\ref{eq:m1}) is the component form of the surface integrals in
\Eq~(\ref{eq:m11}). So the first task of preprocessing is to drive the
raw magnetogram to fulfill the constraints of \Eq~(\ref{eq:m1}) and
thus to be closer to an ideally force-free magnetogram. This task is
also dubbed as `removing force' in the forced magnetogram. The second
task of preprocessing is to smooth the raw data to mimic the field
expansion. Smoothing is also very necessary for the practical
computation based on numerical difference with limited resolution,
which cannot resolve sufficiently small structures in the raw
data. Besides, smoothing can remove measurement noise and increase
signal-to-noise ratio.

%Review of the present method of preprocessing.
Several preprocessing codes \citep{Wiegelmann2006b, Fuhrmann2007,
  Metcalf2008} have been developed and they share the basic approach
proposed by \citet{Wiegelmann2006b}. A functional $L$ is designed by
adding up the $\chi^{2}$ deviations from the constraints of
\Eq~(\ref{eq:m1}), the terms that control deviation from the raw
data and the smoothness with different weights, {\it e.g.},
\begin{equation}
  \label{eq:L}
  L = \mu_{1}L_{1}+\mu_{2}L_{2}+\mu_{3}L_{3}+\mu_{4}L_{4}
\end{equation}
where $\mu$ is the weighting factor, $L_{1} =
F_{x}^{2}+F_{y}^{2}+F_{z}^{2}$, $L_{2} =
T_{x}^{2}+T_{y}^{2}+T_{z}^{2}$, $L_{3}=\int_{S} |\vec B-\vec B_{\rm
  obs}|^{2} \rmd s $, and $L_{4}$ measures the roughness of the
data. Then the target magnetogram is searched by minimizing the
functional $L$ using an optimization method. Different algorithms of
smoothing and optimization have been utilized and the results are more
or less different, as shown in a comparison study by
\citet{Fuhrmann2011}. Also the differences can result from different
choices of the weighting factors.

Under this framework of preprocessing, there are two problems not well
addressed, namely, to what extent the force is needed to be removed
and to what extent the smoothing can be performed? We care about these
problems from both numerical and physical considerations. Ideally we
prefer the map to satisfy the force-free constraints precisely, but
this condition need not be satisfied strictly considering that
numerical discretization error is unavoidable in the extrapolation
with finite resolution. The smoothing also ought not to be done
arbitrarily if we want to mimic the expansion of the field from the
photosphere to some specific height above. Over-smoothing of the data
may smear the basic structures while a too-limited smoothing cannot
filter the small-scale noise sufficiently. A careful choice of the
weighting factors $\mu$ is required to deal with these problems.

This paper is devoted to handling these problems in the
preprocessing. We use the values of force-freeness and smoothness
calculated from numerical potential-field solution at some height
above the photosphere as a reference to guide the preprocessing. Based
on a simple rule that any magnetic field can be split into two parts:
a potential field and a non-potential field, we develop a new
preprocessing code using this splitting of the magnetic field, which
is consistent with our extrapolation code CESE--MHD--NLFFF
\citep{Jiang2012apj,Jiang2012apj1}. We show below how the raw
magnetogram can be driven to force-free and smooth with the same level
as that of the numerical potential field at a height of roughly 400~km
above the photosphere, {\it i.e.}, the bottom of the force-free
domain. The remainder of the paper is organized as follows. In
Section~\ref{sec:method} we give the basic method and formulas, and we
show how to choose the weighting factors in Section~\ref{sec:mu}. We
then apply the method to preprocess two sampled magnetograms taken by
{SDO}/HMI and analyze the results in Section~\ref{sec:res}. Finally
discussion and conclusions are given in Section~\ref{sec:con}.

\section{Method}
\label{sec:method}

Generally the coronal magnetic field can be split into two parts: a
potential field matching the normal component of the bottom
magnetogram, and a non-potential part with the normal field vanishing
at the bottom. Particularly, of the vector magnetogram, the magnetic
field $\vec B$ can be written as
\begin{equation}
  \vec B = \vec B_{0} + \vec B_{1} = (B_{0x}+B_{1x},B_{0y}+B_{1y},B_{0z})
\end{equation}
where $(B_{0x},B_{0y},B_{0z})$ are the components of the potential
part $\vec B_{0}$ and $(B_{1x},B_{1y})$ the components of the
non-potential part $\vec B_{1}$. Note that $B_{0z} = B_{z}$ and
$B_{1z} = 0$.

Supposing $\vec B$ is a force-free magnetogram and since its potential
part $\vec B_{0}$ already fulfills the force-free conditions
of \Eq~(\ref{eq:m1}), we can derive special force-free conditions for its
non-potential part $(B_{1x},B_{1y})$, which are expressed as
\begin{eqnarray}
  \label{eq:m3}
  \int_{S}B_{1x}B_{0z}\rmd x\rmd y = 0,\ \ 
  \int_{S}B_{1y}B_{0z}\rmd x\rmd y = 0,\nonumber\\
  \int_{S}\Gamma_{B}\rmd x\rmd y = 0,\ \
  \int_{S}x\Gamma_{B}\rmd x\rmd y = 0,\ \
  \int_{S}y\Gamma_{B}\rmd x\rmd y = 0,\nonumber\\ 
  \int_{S}(yB_{1x}B_{0z}-xB_{1y}B_{0z}) \rmd x\rmd y = 0
\end{eqnarray}
where we denote $\Gamma_{B} =
B_{1x}^{2}+B_{1y}^{2}+2(B_{0x}B_{1x}+B_{0y}B_{1y})$. The derivation is
straightforward, for example
\begin{eqnarray}
  E_{B} = (B_{0x}+B_{1x})^{2}+(B_{0y}+B_{1y})^{2}-B_{0z}^{2} \nonumber\\
  = E_{B_{0}}+B_{1x}^{2}+B_{1y}^{2}+2(B_{0x}B_{1x}+B_{0y}B_{1y}) =
  E_{B_{0}}+\Gamma_{B},
\end{eqnarray}
and we have
\begin{equation}
  \int_{S}\Gamma_{B}\rmd x\rmd y = \int_{S}E_{B}\rmd x\rmd
  y-\int_{S}E_{B_{0}}\rmd x\rmd y = 0.
\end{equation}
All other expressions in \Eq~(\ref{eq:m3}) can be derived easily in
the similar way.

Let $\mB~(\mB_{x}, \mB_{y}, \mB_{z})$ denote the observed photospheric
field, {\it i.e.}, the raw magnetogram, and its splitting form is 
\begin{equation}
  \mB = \mB_{0} + \mB_{1} = 
  (\mB_{0x}+\mB_{1x},\mB_{0y}+\mB_{1y},\mB_{0z})
\end{equation}
with $\mB_{0}$ and $\mB_{1}$ denoting the potential and non-potential
parts, respectively. Here $\mB_{0}$ is computed based on $\mB_{z}$
using the potential field model and then $\mB_{1}$ is also
obtained. The computation of a potential field needs only the normal
component of the field on the bottom and is now a trivial task, which
can be carried out conveniently by using the Green's function method
\citep{Metcalf2008} or other much faster scheme \citep{Jiang2012SoPh}.
% From a observation point of view, the
% potential part $\mB_{0}$ is more reliable than the non-potential part,
% since $\mB_{0}$ is determined by the observed vertical field which is
% measured much more precisely than the transverse field. 
%*REMOVE This sentence according to the reviewer's COMMENTs*

Generally, $\mB_{1}$ does not fulfill the force-free conditions
of \Eq~(\ref{eq:m3}). If without smoothing, we only need to let $\vec
B_{0} = \mB_{0}$ and reduce the non-potential part $\mB_{1}$ to
$\vec B_{1}$ satisfying \Eq~(\ref{eq:m3}). For the purpose of
smoothing, $\vec B_{0}$ is obtained by taking the
data at a plane just one pixel above the photosphere from the 3D
potential field extrapolated from the observed $\mB_{0z}$. This is
suitable for the {SDO}/HMI data which has a pixel size of about 360~km
({\it i.e.}, 0.5 arcsec), an approximate height above which the coronal
field becomes force-free according to \citet{Metcalf1995}.
% --REVIEW
For magnetograms with other sizes of pixel, we need to take the
potential field data at a given physical height (where the force-free
assumption becomes valid, {\it e.g.}, 400~km) and not necessarily one pixel
above the photosphere.
% --
$\vec B_{0}$ obtained in this way can be regarded as the potential
part of the chromospheric field, also a preprocessed
counterpart of $\mB_{0}$.
% --REVIEW
Of course, if the measurements of chromospheric longitudinal fields
are available \citep[{\it e.g.},][]{Yamamoto2012}, we recommend using those
data directly to construct the potential part $\vec B_{0}$, which is
certainly preferred over that based on the photospheric
$\mB_{0z}$.

The second task, to reduce $\mB_{1}$ to $\vec B_{1}$, is carried out
using an optimization method similarly to \citet{Wiegelmann2006b}. We
intend to minimize the total magnetic force and torque which are
quantified by
\begin{eqnarray}
  L_{1} = L_{11}^{2}+L_{12}^{2}+L_{13}^{2},\ \ 
  L_{2} = L_{21}^{2}+L_{22}^{2}+L_{23}^{2}
\end{eqnarray}
where for convenience of presentation we denote
\begin{eqnarray}
  L_{11} &\equiv& \sum_{\rm p}B_{1x}B_{0z},\ \ 
  L_{12} \equiv \sum_{\rm p}B_{1y}B_{0z},\ \ 
  L_{13} \equiv \sum_{\rm p}\Gamma_{B},\\
  L_{21} &\equiv& \sum_{\rm p}x\Gamma_{B},\ \
  L_{22} \equiv \sum_{\rm p}y\Gamma_{B},\ \ 
  L_{23} \equiv \sum_{\rm p}(yB_{1x}B_{0z}-xB_{1y}B_{0z}).
\end{eqnarray}
Here the summation $\sum_{\rm p}$ is over all the pixels of the
magnetogram, and these summations are the numerical counterparts of
the integrals in \Eq~(\ref{eq:m3}). 

The observation term $L_{3}$ (to restrict the deviation from the
observed data) and smoothing functional $L_{4}$ (to control the
smoothness) are also considered by \citet{Wiegelmann2006b}
\begin{eqnarray}
  L_{3} &=&  
  \sum_{\rm p}\left[(B_{1x}-\mB_{1x})^{2}+ 
    (B_{1y}-\mB_{1y})^{2}\right],
  \nonumber\\
  L_{4} &=& \sum_{\rm p}\left[(\Delta B_{1x})^{2}+
    (\Delta B_{1y})^{2}\right]
\end{eqnarray}
where $\Delta$ is a usual five-point 2D-Laplace operator, {\it i.e.}, for
the pixel $(i,j)$
\begin{equation}
  \Delta B_{i,j} \equiv B_{i+1,j}+B_{i-1,j}+B_{i,j+1}+B_{i,j-1}-4B_{i,j}.
\end{equation}
This simply states that the smaller $L_{4}$ gives the smoother data.

Additionally, the above functionals $L_{\ell}$ (where $\ell =
1,2,3,4$) are normalized by $N_{\ell}$ which are given by
\begin{eqnarray}
  N_{1} &=& \left(\sum_{\rm p}|\mB|^{2}\right)^{2},\ \ 
  N_{2} = \left(\sum_{\rm p}\sqrt{x^{2}+y^{2}}|\mB|^{2}\right)^{2},
  \nonumber\\
  N_{3} &=& \sum_{\rm p}(\mB_{x}^{2}+\mB_{y}^{2}),\ \ 
  N_{4} = \sum_{\rm p}\left[\left(\overline{\Delta}\mB_{1x}\right)^{2}
    +\left(\overline{\Delta}\mB_{1y}\right)^{2}\right]
\end{eqnarray}
where
\begin{equation}
  \overline{\Delta}B_{i,j} \equiv B_{i+1,j}+B_{i-1,j}+B_{i,j+1}+B_{i,j-1}+4B_{i,j}.
\end{equation}

We use a steepest descent method \citep{Press1992} to minimize a
weighted average of the above functionals
\begin{equation}
  \label{eq:L}
  L=\sum_{\ell =1}^{4}\frac{\mu_{\ell}}{N_{\ell}}L_{\ell}
\end{equation}
where $\mu_{\ell}$ is the weighting factor. Generally, the weighting
factors for the magnetic force and torque are simply given by
$\mu_{1}=\mu_{2}=1$ since there is no obvious reason to give bias on
any of these two quantities. The determination of $\mu_{3}$ and
$\mu_{4}$ will be described in the next section, and different
combinations of them are tested for two HMI magnetograms to search the
optimal choice in Section~\ref{sec:res}.

Since $L$ is an explicit functional of the arguments
$(B_{1x},B_{1y})$, its gradient $\nabla L$ can be expressed at each
pixel $q$ as
\begin{eqnarray}
  \frac{\partial L}{\partial (B_{1x})_{q}} = 
  2\frac{\mu_{1}}{N_{1}}\left[
    L_{11}(B_{0z})_{q}+
    L_{13}(2B_{1x}+2B_{0x})_{q}\right]
  \nonumber\\
  +2\frac{\mu_{2}}{N_{2}}\left[
    L_{21}(2xB_{1x}+2xB_{0x})_{q}+
    L_{22}(2yB_{1x}+2yB_{0x})_{q}+
    L_{23}(yB_{0z})_{q}
  \right]
  \nonumber\\
  +2\frac{\mu_{3}}{N_{3}}(B_{1x}-\mB_{1x})_{q}
  +2\frac{\mu_{4}}{N_{4}}(\Delta(\Delta B_{1x}))_{q},\\
  \frac{\partial L}{\partial (B_{1y})_{q}} = 
  2\frac{\mu_{1}}{N_{1}}\left[
    L_{12}(B_{0z})_{q}+
    L_{13}(2B_{1y}+2B_{0y})_{q}\right]
  \nonumber\\
  +2\frac{\mu_{2}}{N_{2}}\left[
    L_{21}(2xB_{1y}+2xB_{0y})_{q}+
    L_{22}(2yB_{1y}+2yB_{0y})_{q}+
    L_{23}(-xB_{0z})_{q}
  \right]
  \nonumber\\
  +2\frac{\mu_{3}}{N_{3}}(B_{1y}-\mB_{1y})_{q}
  +2\frac{\mu_{4}}{N_{4}}(\Delta(\Delta B_{1y}))_{q}.
\end{eqnarray}
The procedure of the steepest descent is performed as
follows. We start from an initial guess $(B_{1x}^{0},B_{1y}^{0})$,
{\it e.g.}, the observed data $(\mB_{1x}, \mB_{1y})$, and
march the solution in each iteration $k$ along the steepest
descent direction ({\it i.e.}, opposite to the gradient direction) by
\begin{equation}
  (B_{1x}^{k+1})_{q} = (B_{1x}^{k})_q-
  \lambda_{k}\frac{\partial L}{\partial (B_{1x}^{k})_{q}},\ \ 
  (B_{1y}^{k+1})_{q} = (B_{1y}^{k})_q-
  \lambda_{k}\frac{\partial L}{\partial (B_{1y}^{k})_{q}}.
\end{equation}
It is important to choose a proper step size $\lambda_{k}$ at each
step to maximize the local descent. This can be determined by a
bisection line-search algorithm to solve the one-dimensional
optimization problem at every iteration
\begin{equation}
  \lambda_{k} = \arg\min L(\vec B_{1}^{k}-\lambda \nabla L(\vec B_{1}^{k})).
\end{equation}
The iteration is terminated if the maximum residual of the field $[{\rm
res}(\vec B_{1})]_{\max}$, defined by
\begin{equation}
  \label{res}
  [{\rm res}(\vec B_{1})]_{\max} = \lambda_{k} \max[\nabla L(\vec B_{1}^{k}) ],
\end{equation}
is smaller than $0.1$~gauss (G) for 10 successive steps.

\section{Choice of the Weighting Factor $\mu$}
\label{sec:mu}
%How to choose the weighting factor $\mu$

A careful choice of optimal weighting factors $\mu$ is critical for a
good performance of preprocessing. Using the qualities of
force-freeness and smoothness of the numerical potential part $\vec
B_{0}$ as a reference,
%Choosing the weighting factors $\mu_{3}$ and $\mu_{4}$ is made 
we guide the optimization of $\mu$ according to the follow
constraints:

%\begin{enumerate}
i) The residual force and torque in the target magnetogram $\vec B$
should be reduced to the same order of those in $\vec B_{0}$;

ii) The smoothness of the target magnetogram $\vec B$ should reach the
same level as that of $\vec B_{0}$ (Since $B_{z} = B_{0z}$, it means
that the smoothness of $B_{x}$ and $B_{y}$ should match that of
$B_{z}$. This is reasonable since there is no preference for any
component of the vector);

iii) With the constraints i) and ii) fulfilled, the deviation
between the target magnetogram $\vec B$ and the observed data $\mB$
should be minimized.
%\end{enumerate}

In the constraints the residual magnetic force and torque of the data
are quantified by two parameters, $\epsilon_{\rm force}$ and
$\epsilon_{\rm torque}$, defined as usual
\begin{eqnarray}
  \epsilon_{\rm force} &=& \frac{|\sum_{\rm p} B_{x}B_{z}|+|\sum_{\rm p}
    B_{y}B_{z}|+|\sum_{\rm p} E_{B}|}{\sum_{\rm p} (B_{x}^{2}+B_{y}^{2}+B_{z}^{2})},\\
  \epsilon_{\rm torque} &=& \frac{|\sum_{\rm p} xE_{B}|+|\sum_{\rm p} yE_{B}|+|\sum_{\rm p} (
    yB_{x}B_{z}-xB_{y}B_{z})|}{\sum_{\rm p} \sqrt{x^{2}+y^{2}}(B_{x}^{2}+B_{y}^{2}+B_{z}^{2})}.
\end{eqnarray} 
and the smoothness of component $B_{m}$ ($m=x,y,z$) is measured by
\begin{equation}
  S_{m} = \sum_{\rm p}\left[(\Delta B_{m})^{2}\right]/
  \sum_{\rm p}\left[(\overline{\Delta} B_{m})^{2}\right].
\end{equation}

Beside the above constraints the total iteration steps needed by
computation is also considered if the magnetogram's resolution is
very high, since the computing time of the preprocessing may be rather
long.

%++++++++++++++++++++++++++++++++++++++++++++++++++++++++++++
\begin{figure}[htbp]
  \centering
  \includegraphics[width=0.48\textwidth]{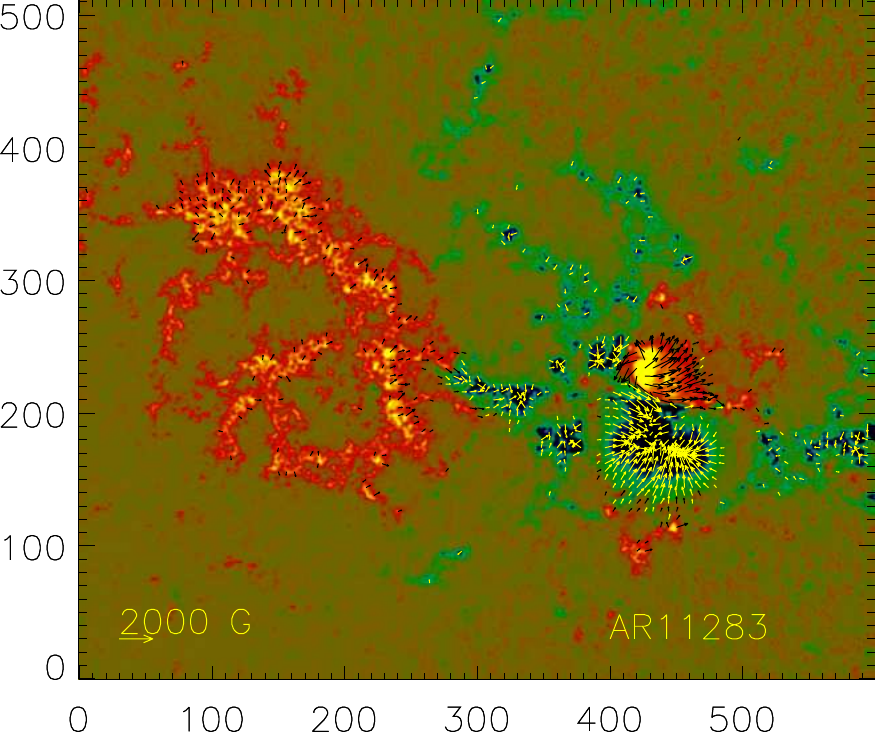}
  \includegraphics[width=0.48\textwidth]{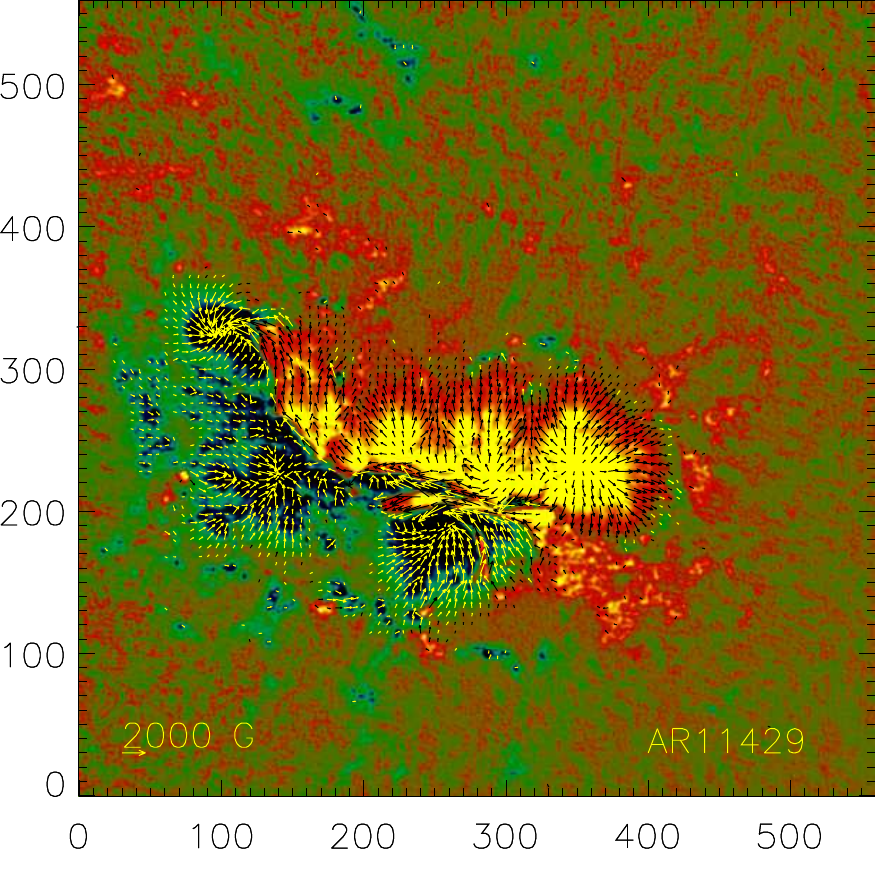}
  \caption{The observed vector magnetograms for AR 11283 at 05:36 UT
    on 8 September 2011 and AR 11429 at 00:00~UT on 7 March 2012. The
    background shows the vertical components with saturation values of
    $\pm 1000$~G; the vectors represent the transverse field and only
    the field stronger than $200$~G is plotted. The length unit is
    0.5~arcsec.}
  \label{fig:rawmaps}
\end{figure}

\begin{figure}[htbp]
  \centering
  \includegraphics[width=\textwidth]{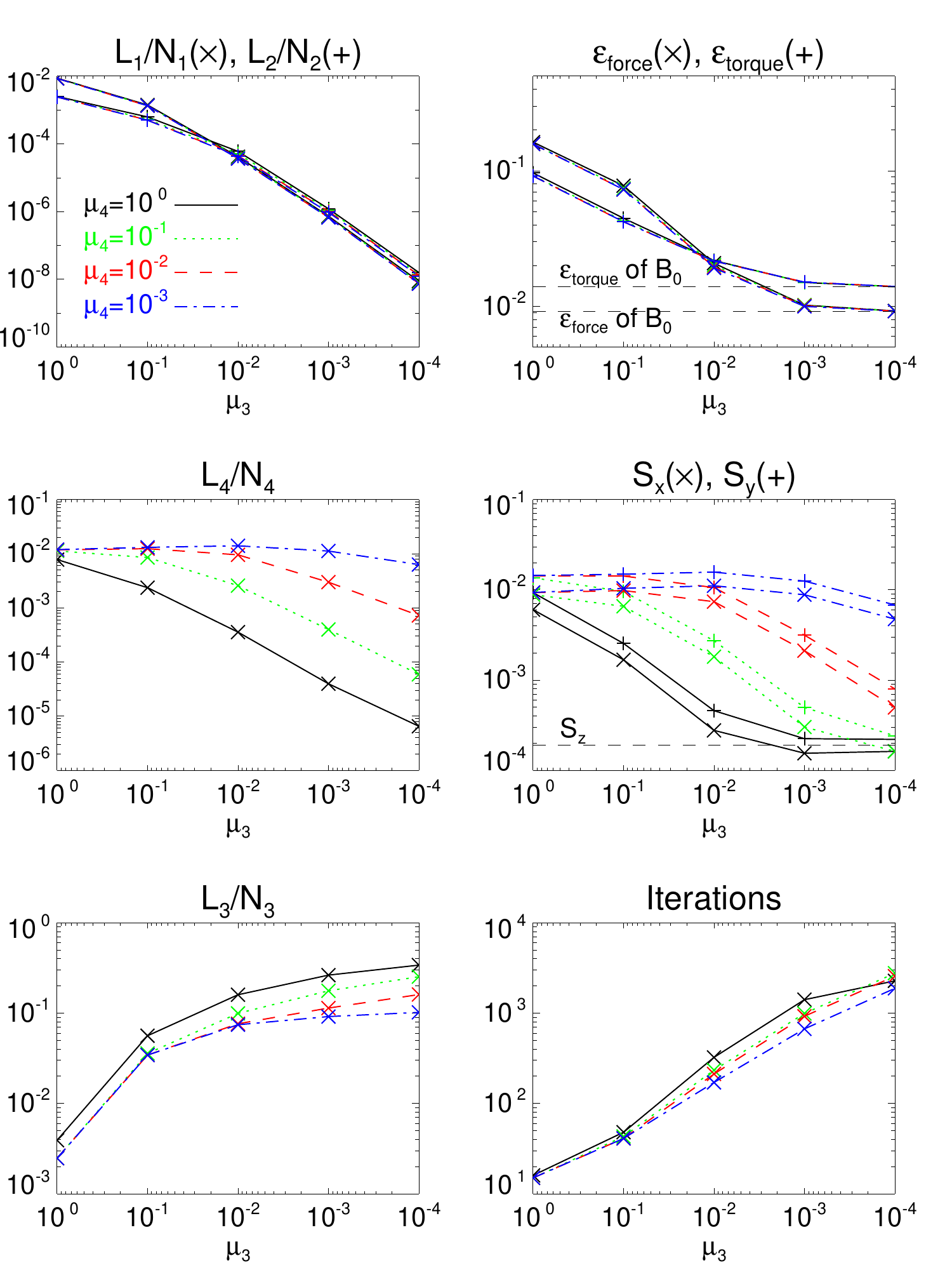}
  \caption{Preprocessed results for the magnetogram of AR 11283 with
    different $\mu_{3}$ and $\mu_{4}$. Results for different values of
    $\mu_{4}$ are plotted using different colors and line styles as
    denoted in the top left panel. The dashed lines in the top right
    and middle right panels represent the parameter values of the
    potential part $\vec B_{0}$, which is used as the reference to
    choose the optimal weights. The optimal weights are chosen such
    that $\epsilon_{\rm force},\epsilon_{\rm torque}$, $S_{x}$, and
    $S_{y}$ are close to those of $\vec B_{0}$ and $L_{3}/N_{3}$ is
    minimized. }
  \label{fig:result11283}
\end{figure}

\begin{figure}[htbp]
  \centering
  \includegraphics[width=\textwidth]{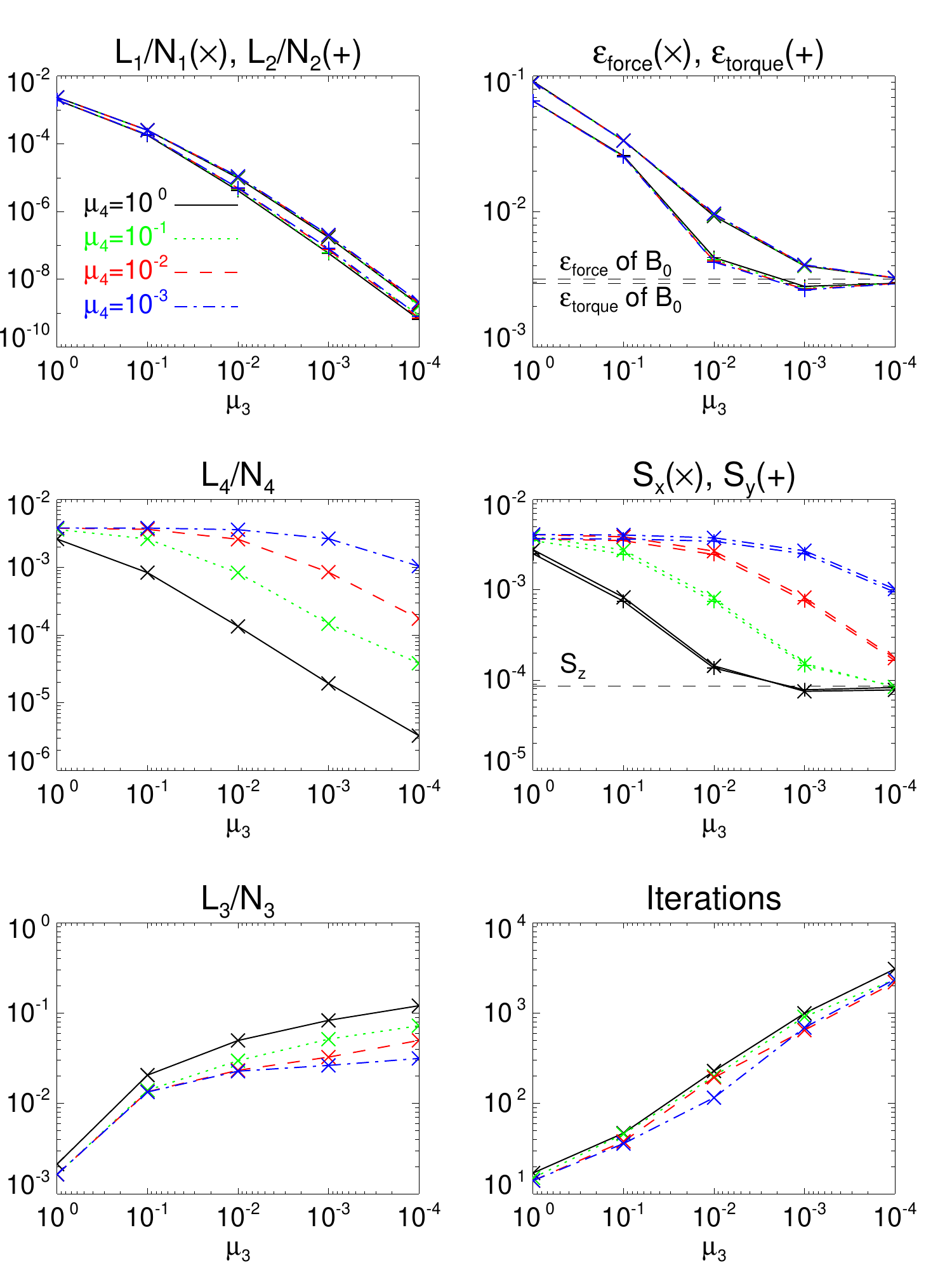}
  \caption{Same as \Fig~\ref{fig:result11283} but for AR 11429.}
  \label{fig:result11429}
\end{figure}

\begin{table}[htbp]
  \caption{Quality of the magnetograms. The preprocessed results are
    produced with the optimal weighting factors $\mu_{3}=0.001$ and
    $\mu_{4}=1$. The parameter $\epsilon_{\rm flux}$ is the
    total magnetic flux normalized by the total unsigned flux.}
  \centering
  \begin{tabular}{lllllll}
    \hline
    Data & $\epsilon_{\rm flux}$ & $\epsilon_{\rm force}$ & $\epsilon_{\rm torque}$ & $S_{x}$ & $S_{y}$ & $S_{z}$\\
    \hline
    AR 11283\\
    Raw                 & -7.88E-02 & 2.84E-01 & 2.38E-01 &8.38E-03 & 1.28E-02 & 2.49E-03 \\
    Preprocessed        & -8.98E-02 & 1.02E-02 & 1.50E-02 &1.55E-04 & 2.25E-04 & 1.92E-04\\
    Numerical potential & -8.98E-02 & 9.14E-03 & 1.40E-02 &2.10E-04 & 1.78E-04 & 1.92E-04\\
    \hline
    AR 11429\\
    Raw map          & -1.36E-02 & 1.82E-01 & 1.55E-01 & 3.76E-03 & 3.43E-03 & 1.21E-03 \\
    Preprocessed map & -1.46E-02 & 3.98E-03 & 2.81E-03 & 7.52E-05 & 7.82E-05 & 8.61E-05 \\
    Numerical potential & -1.46E-02 & 3.19E-03 & 2.96E-03 &8.38E-05 &8.53E-05 & 8.61E-05\\
    \hline
  \end{tabular}
  \label{tab:quality}
\end{table}

\begin{figure}[htbp]
  \centering
  \includegraphics[width=\textwidth]{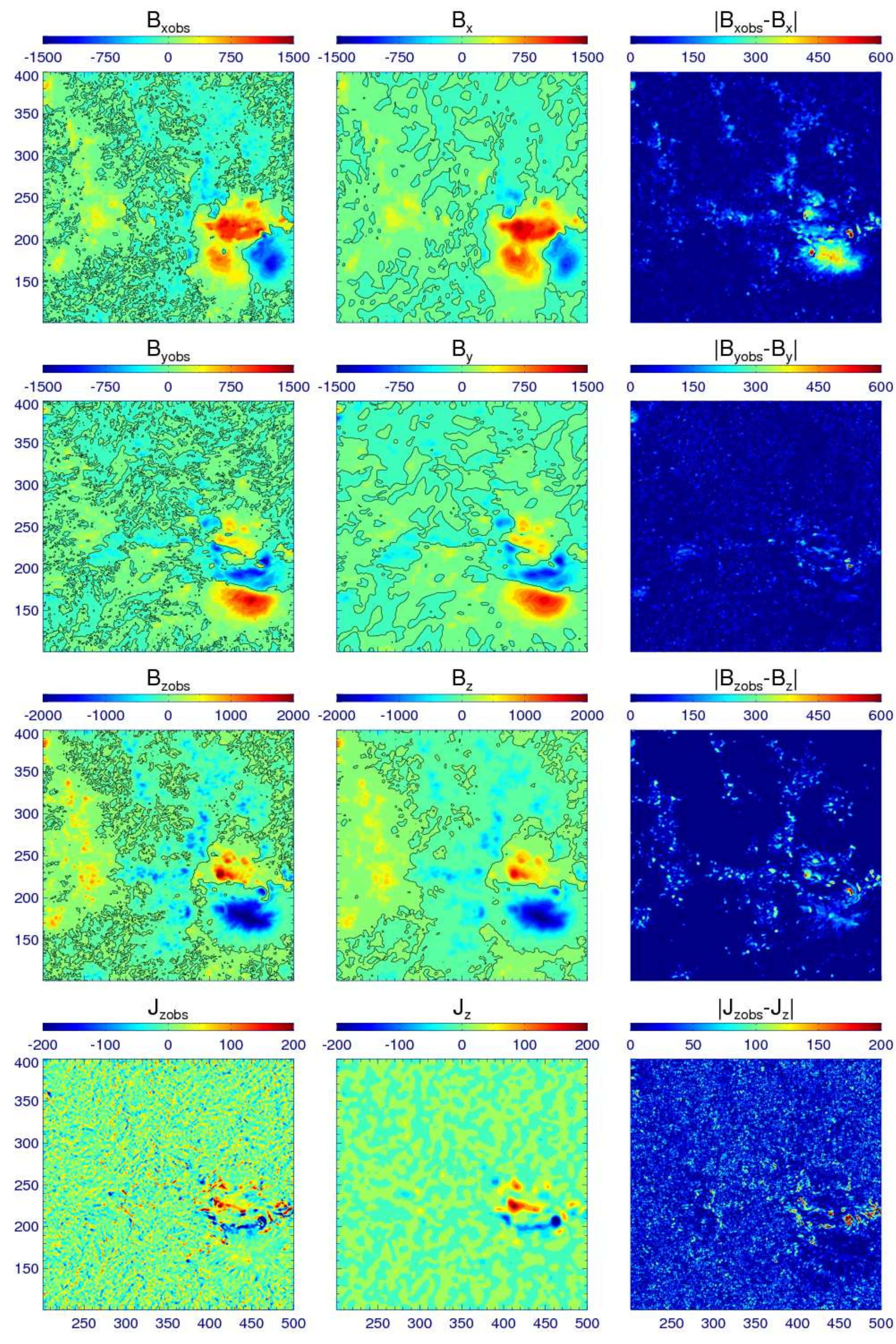}
  \caption{Comparison of the raw magnetogram and preprocessed
    magnetogram for AR 11283. The left column is the raw data, the
    middle column is the preprocessed data and the right column is the
    absolute differences between them. Rows from top to bottom are the
    three components of the data and the vertical current $J_{z}$,
    respectively. Over the images of $B_{x},B_{y}$, and $B_{z}$
    components are the contour lines of their zero values.}
  \label{fig:map11238}
\end{figure}

\begin{figure}[htbp]
  \centering
  \includegraphics[width=\textwidth]{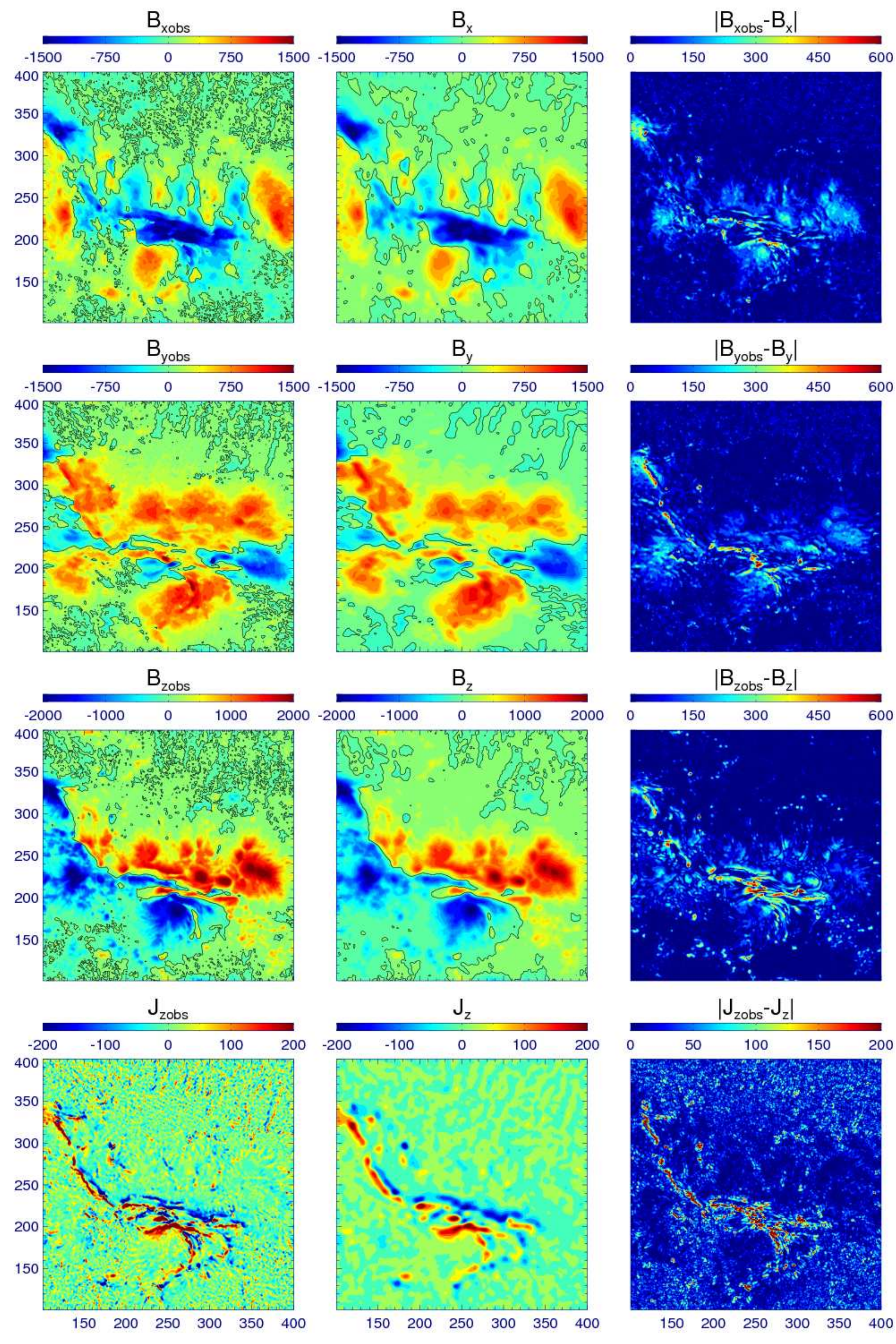}
  \caption{Same as \Fig~\ref{fig:map11238} but for AR 11429.}
  \label{fig:map11429}
\end{figure}

\begin{figure}[htbp]
  \centering
  \includegraphics[width=0.48\textwidth]{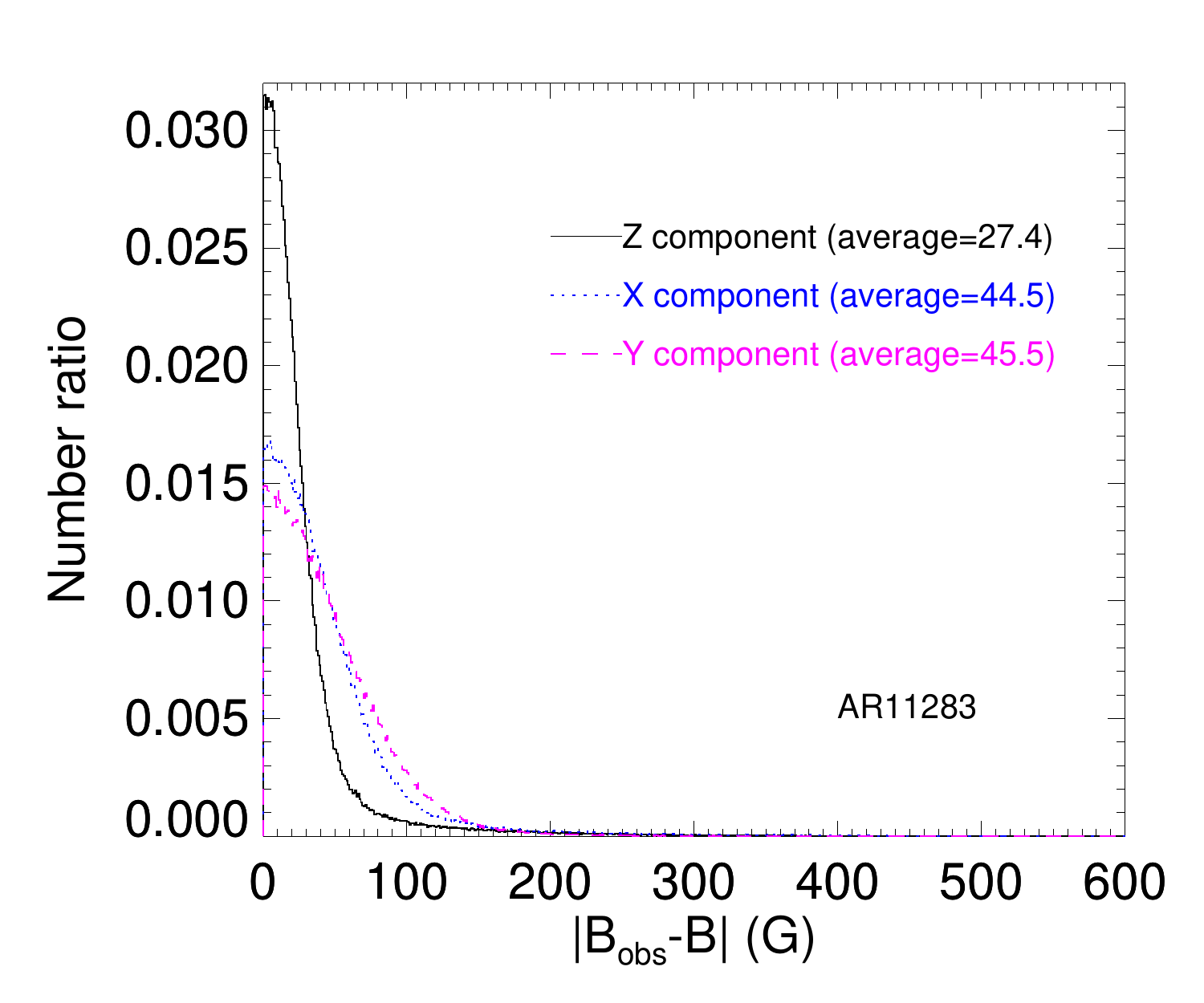}
  \includegraphics[width=0.48\textwidth]{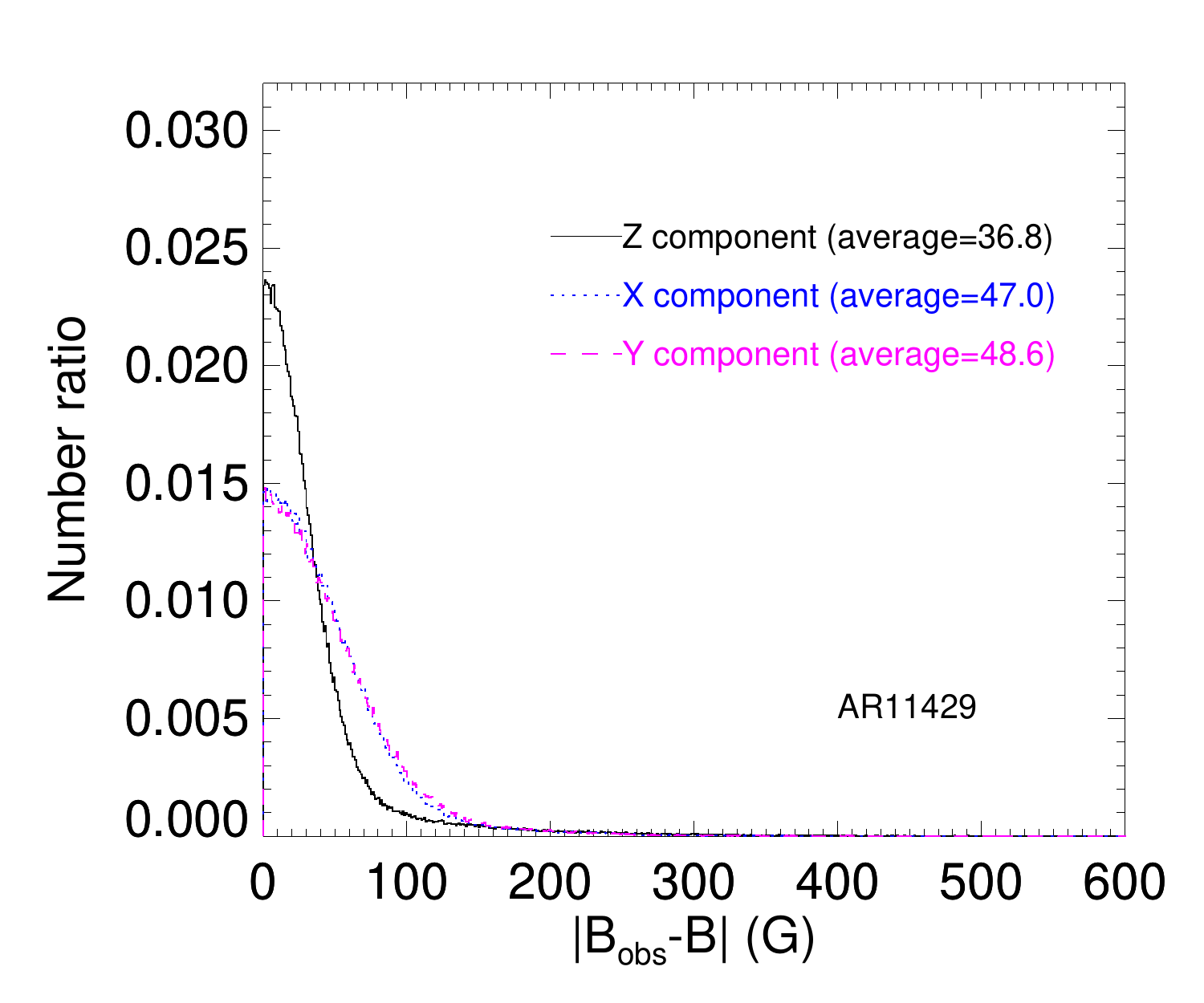}
  \caption{Histograms of the changes of the vector components for all
    the pixels between the raw and preprocessed magnetograms. The
    horizontal axis represents the the absolute values of the
    differences between the raw and preprocessed data, and the
    vertical axis represents the number of the pixels normalized by
    the total number. The average changes are also labelled on the
    figure.}
  \label{fig:dBhistrogram}
\end{figure}

\begin{figure}[htbp]
  \centering
  \includegraphics[width=\textwidth]{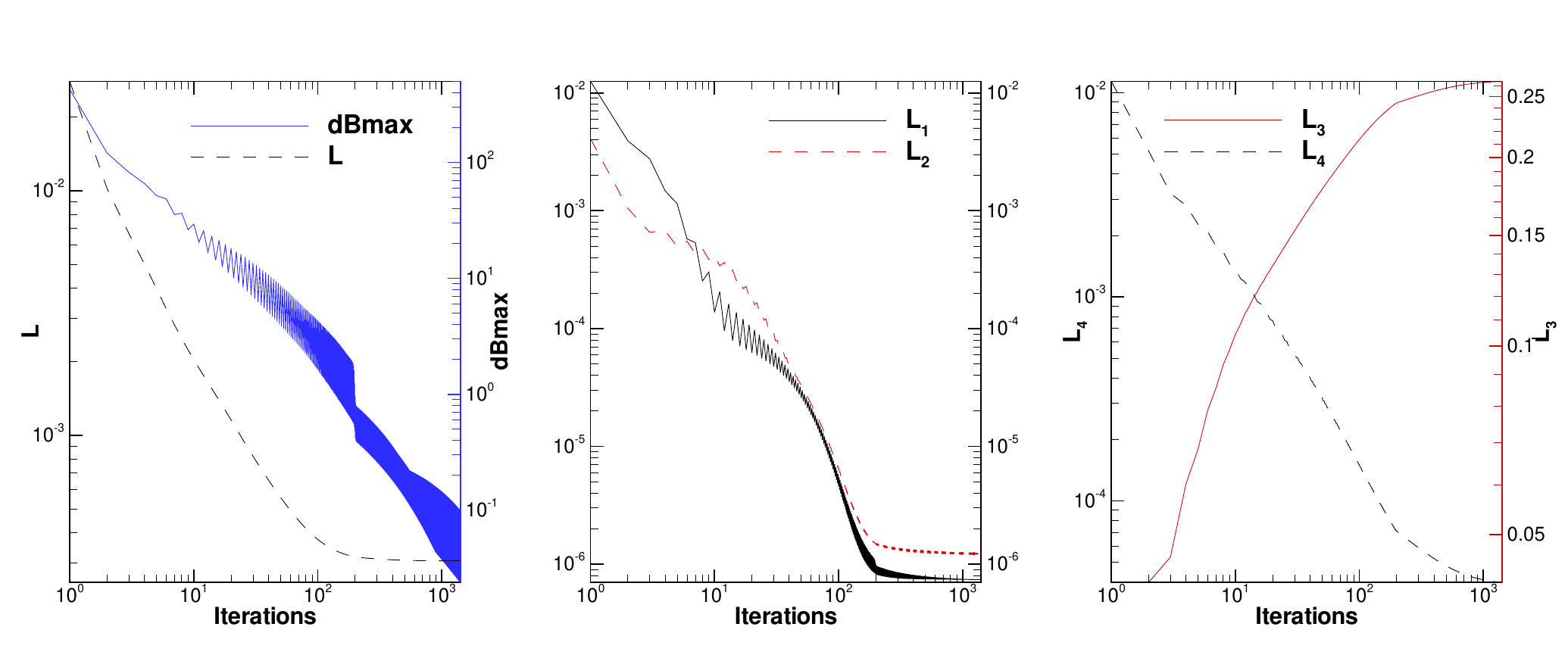}
  \caption{Evolutions of the functional with iterations in the
    optimization process. dBmax is the maximum residual of the field
    in each iteration step, see \Eq~(\ref{res}).}
  \label{fig:iter}
\end{figure}

%++++++++++++++++++++++++++++++++++++++++++++++++++++++++++++

\section{Preprocessing the {SDO}/HMI Magnetograms}
\label{sec:res}

In this section we apply the preprocessing code to several
magnetograms taken by {SDO}/HMI, and search the optimal values for the
weighting factors. The {\it Helioseismic and Magnetic Imager} (HMI) on
board the {\it Solar Dynamics Observatory} (SDO) provides photospheric
vector magnetograms with a high resolution both in space and time. It
observes the full Sun with a 4k$\times$4k CCD whose spatial sampling
is 0.5 arcsec per pixel. Raw filtergrams are obtained at six different
wavelengths and six polarization states in the Fe {\sc i} 6173~{\AA}
absorption line, and are collected and converted to observable
quantities (like Dopplergrams, continuum filtergrams, and
line-of-sight and vector magnetograms) on a rapid time cadence. For
the vector magnetic data, each set of filtergrams takes 135 s to be
completed. To obtain vector magnetograms, Stokes parameters are first
derived from filtergrams observed over a 12-min interval and then
inverted through the Very Fast Inversion of the Stokes Vector
\citep{Borrero2011}. The 180$^{\circ}$ azimuthal ambiguity in the
transverse field is resolved by an improved version of the ``minimum
energy'' algorithm \citep{Leka2009}. Regions of interest with strong
magnetic field are automatically identified near real time
\citep{Turmon2010}. A detailed description on how the vector
magnetograms are produced can be found on the website
\url{//http://jsoc.stanford.edu/jsocwiki/VectorPaper}.

The raw magnetograms we use here were downloaded from
\url{http://jsoc.stanford.edu/jsocwiki/ReleaseNotes2}, where the HMI
vector magnetic field data series \texttt{hmi.B\_720s\_e15w1332} are
released for several active regions. There are two special formats,
{\it i.e.}, direct cutouts and remapped images. We use the remapped
format which is more suitable for modeling in local Cartesian
coordinates, since the images are computed with a Lambert cylindrical
equal area projection centered on the tracked region. For our test, we
select two active regions, AR 11283 and AR 11429, both of which
produced X-class flares and thus were very
non-potential. \Fig~\ref{fig:rawmaps} shows the magnetograms for AR
11283 at 05:36 UT on 8 September 2011 and AR 11429 at 00:00~UT on 7
March 2012. The size of the magnetograms are respectively $600\times
512$ and $560\times 560$ pixels.

In \Figs~\ref{fig:result11283} and \ref{fig:result11429} we show the
preprocessing results with different sets of $\mu_{3}$ and
$\mu_{4}$. Since for such large magnetograms it is nontrivial to
perform plenty of tests with continuous sets of weighting factors to
pick the optimal one, we only compute the results for several groups
of $\mu_{3}$ and $\mu_{4}$ as shown in the figures, {\it i.e.},
$\mu_{3}=1, 0.1, 0.01, 0.001, 0.0001$ and $\mu_{4} = 1, 0.1, 0.01,
0.001$. For each set of weighting factors, the normalized terms
$L_{l}/N_{l}$ and the smoothness and the force-free quality
$\epsilon_{\rm force}$ and $\epsilon_{\rm torque}$ are plotted. By
comparing the results with fixed $\mu_{3}$ but different $\mu_{4}$, we
can see that the force-free parameters
$L_{1}/N_{1},L_{2}/N_{2},\epsilon_{\rm force}$, and $\epsilon_{\rm
  torque}$ are almost entirely determined by $\mu_{3}$. When
decreasing $\mu_{3}$, {\it i.e.}, allowing more freedom of modifying
the raw data, $L_{1}$ and $L_{2}$ decrease very quickly (their
magnitude decreases quicker than that of $\mu_{3}$), but the residual
force parameters $\epsilon_{\rm force}$ and $\epsilon_{\rm torque}$
reach a minimum and cannot be reduced any further. This is because the
potential part $\vec B_{0}$ has a non-zero value of $\epsilon_{\rm
  force}$ and $\epsilon_{\rm torque}$ (due to numerical error of
finite resolution), which is the minimum of $\epsilon_{\rm
  force},\epsilon_{\rm torque}$ that can be reached for the target
magnetograms. The results for both tests show that $\mu_{3}=0.001$ is
small enough which gives $\epsilon_{\rm force}$ and $\epsilon_{\rm torque}$
nearly the same as those of $\vec B_{0}$, meaning that the force in
the non-potential part $\vec B_{1}$ is decreased near or under the
level of numerical error. Even smaller $\mu_{3}$ cannot improve the
quality of force-freeness, but deviates the target magnetograms
farther away from the original data. Thus we set the optimal value of
$\mu_{3}=0.001$. With a given value of $\mu_{3}$, the values of
smoothness are controlled by $\mu_{4}$. Obviously $\mu_{4} = 1$ is a
good choice which gives the values of $S_{x}$ and $S_{y}$ very similar
to $S_{z}$, meaning that the smoothness of the target magnetograms is
consistent with their potential part $\vec B_{0}$. We believe the
choice of weight for smoothness here is more physics-based than in
other methods in which it is given more or less arbitrarily.

The results of preprocessing the two magnetograms with the optimal
weighting factors $\mu_{3}=0.001$ and $\mu_{4}=1$ are summarized in
Table~\ref{tab:quality}. \Figs~\ref{fig:map11238} and
\ref{fig:map11429} show a comparison of the original and preprocessed
magnetograms. Here the color-map is designed to manifest both strong
and weak fields. As shown, the map after preprocessing resamples the
feature of the original data while small structures tend to be
smoothed out. In the bottom of the figures we plot the results for the
vertical current $J_{z}$ which is calculated by taking finite
differences of the transverse field
\begin{equation}
  \label{eq:alpha}
  J_{z}^{i,j} = \frac{B_{y}^{i+1,j}-B_{y}^{i-1,j}}{2}
  -\frac{B_{x}^{i,j+1}-B_{x}^{i,j-1}}{2}.
\end{equation}
Since the numerical differences are very sensitive to noise, random
noise in the data exhibits more clearly in the $J_{z}$ map,
particularly in weak field regions, and they are suppressed
effectively by the smoothing. Histograms are plotted in
\Fig~\ref{fig:dBhistrogram} to show the distributions of the absolute
differences between the raw and preprocessed magnetograms over all the
pixels. Apparently different approaches of modifying the data give
different distributions, {\it i.e.}, the distribution for the
$z$-component is distinct from those for the $x$ and
$y$-components. This is because the modification for $B_{z}$ is
determined solely by the potential model, while modifications for
$B_{x},B_{y}$ are additionally made by the optimization process. The
change in the vertical field is less than those for the transverse
field. This is consistent with the observation which measures the
line-of-sight component much more precisely than the transverse field,
and thus we have more freedom to modify the transverse field.
%--Reviewer's comments
Still it should be noted that here the potential-field modeled $B_{z}$
may not approximate well the real chromospheric $B_{z}$, which is
preferred to be taken by direct measurements if available.

We finally show in \Fig~\ref{fig:iter} the process of iteration in the
optimization method. Only the result for AR 11283 is plotted as an
example. As shown, the functional $L$ decreased very quickly. By only
200 iteration steps, it almost reaches its minimum, reduced by about
two orders of magnitude from its initial value. The sub-functionals
$L_{1}$ and $L_{2}$ have similar evolution, although with small
oscillations, and the sub-functional $L_{4}$ keeps decreasing but very
slowly after 200 steps.

\section{Conclusions}
\label{sec:con}

% summary of the method
In this paper we have developed a new code of preprocessing the
photospheric vector magnetograms for NLFFF extrapolation. The method
is based on a simple rule that any vector magnetograms can be split
into a potential field part and a non-potential part and we deal with
two parts separately. Preprocessing of the potential part is simply
performed by taking the data sliced at a plane about $400$~km above
the photosphere from the 3D potential-field numerical solution, which
is extrapolated from the observed vertical field. Then the
non-potential part is modified by an optimization method to fulfill
the constraints of total magnetic force-freeness and
torque-freeness. As for practical computation based on numerical
discretization, a strict satisfaction of force-free constraints is
apparently not necessary. Also the extent of the smoothing to be
applied to the data need to be carefully determined, if we want to
mimic the field expansion from the photosphere to some specific height
above. We use the values of force-freeness and smoothness calculated
from the preprocessed potential-field part as a reference to guide the
preprocessing of the non-potential field part, {\it i.e.}, we require
that the target magnetograms have the same level of force-freeness and
smoothness as the reference data. These requirements can restrict well
the values of the free parameters, {\it i.e.}, the weighting factors
in the optimization functional. The code is applied to the {SDO}/HMI
data. Preprocessed results show that the method can remove efficiently
the force and noise, if we properly choose the weighting factors. For
two sampled HMI magnetograms, we find that the optimal weights are
$\mu_{3}=0.001$ and $\mu_{4}=1$, with which the target magnetgrams can
be driven to force-free and smooth with the same level as that of the
numerical potential field at the bottom of the force-free domain.

% how about the results for the extrapolation
The preprocessing code here is developed as a sub-program for a
project of applying our extrapolation code CESE--MHD--NLFFF
\citep{Jiang2012apj1,Jiang2012apj} to observed data. We have applied
CESE--MHD--NLFFF to {SDO}/HMI data with both raw and preprocessed
magnetograms \citep{Jiang2013NLFFF}. By a careful comparison of
the results, we find that the quality of extrapolation is indeed
improved using the preprocessed magnetograms, including the
force-freeness of the results ({\it e.g.}, measured by a
current-weighted mean angle between the magnetic field $\vec B$ and
electric current $\vec J$) and the free energy contents. For example
in the extrapolation of AR 11283, the mean angle between $\vec B$ and
$\vec J$ for the entire extrapolation box of $600\times 512\times 300$
pixels is reduced from $24^{\circ}$ to $17^{\circ}$ and the free
energy is increased from $\approx 0.5\times 10^{32}$~erg to $1.0\times
10^{32}$~erg.

\begin{acks}
  This work is jointly supported by the 973 program under grant
  2012CB825601, the Chinese Academy of Sciences (KZZD-EW-01-4), the
  National Natural Science Foundation of China (41204126, 41274192,
  41031066, and 41074122), and the Specialized Research Fund for State
  Key Laboratories. Data are courtesy of NASA/{SDO} and the HMI
  science teams. The authors thank the anonymous referee for
  invaluable comments.
\end{acks}

\end{article}

\end{document}